\documentclass[11pt]{article}
\usepackage[preprint]{acl}
\usepackage{times}
\usepackage[T1]{fontenc}
\usepackage{latexsym}
\usepackage{amsmath}
\usepackage{amsfonts}
\usepackage{amssymb}
\usepackage{booktabs}
\usepackage{multirow}
\usepackage{graphicx}
\usepackage{subcaption}
\usepackage{algorithm}
\usepackage{algpseudocode}
\usepackage{microtype}

\title{S2A3: Thompson Sampling and Stochastic Exposure Control for High-Stakes CATs}

\author{James Sharpnack \\
  Duolingo \\\And
  Alexander Tsigler \\
  Duolingo \\\And
  J.~R. Lockwood \\
  Duolingo \\\And
  Steven Nydick \\
  Duolingo \\\And
  Alina A. von Davier \\
  Duolingo}

\begin{document}
\maketitle

\begin{abstract}
High-stakes computerized adaptive tests (CATs) require a continuous supply
of calibrated items, yet traditional item piloting is slow, expensive, and
operationally hazardous.
We introduce the S2A3 framework---Soft Scoring (S2) and Adaptive Adaptive
Administration (A3)---which unifies item calibration and test administration
into a single online process.
Thompson sampling enhances item selection by drawing provisional parameters
from each item's posterior distribution and selecting the item maximizing
expected Fisher information, naturally routing uncertain items to informative
test-takers while maintaining measurement precision.
Soft scoring integrates over parameter uncertainty so that incompletely
calibrated items exert appropriately attenuated influence on ability estimates.
A stochastic variant of Sympson-Hetter exposure control balances measurement
efficiency against bank security via a tunable temperature parameter and
item-specific weights.
We validate S2A3 on Yes/No Vocabulary and Vocabulary-in-Context tasks from
the Duolingo English Test, demonstrating rapid item calibration and
preserved scoring reliability even when cold-start items constitute a
significant fraction of the active pool.
\end{abstract}

\section{Introduction}

In a high-stakes digital assessment, it is important to continually grow and
improve the item bank by adding new items and cycling out poorly performing
ones.
New items need to be calibrated prior to being operationalized, yet item
calibration typically requires test-taker responses to fit an item response
theory (IRT) model.
These responses are often gathered in a separate piloting phase, where
test-takers either answer items that do not contribute to their score, or in a
lower-stakes practice test.
Both of these practices are problematic because they can waste test-taker
effort, provide biased response data, or raise test security issues by exposing
operational items \citep{Sharpnack2024}.

This paper presents S2A3, a unified, fully Bayesian framework for
administration and scoring of high-stakes CATs that enables new items to be
launched directly on the operational test without a separate pilot phase.
Its two interlocking components are:
\textbf{Soft Scoring (S2)}, which integrates over item parameter uncertainty
so that incompletely calibrated items exert appropriately reduced influence on
ability estimates; and
\textbf{Adaptive Adaptive Administration (A3)}, which applies Thompson
sampling to route uncertain items to the test-takers who will be most
informative for calibrating them, combined with stochastic Sympson-Hetter
exposure control to prevent any item from dominating the bank.

Together, these pillars address the cold-start problem in CAT:
new items enter the operational pool with parameter distributions predicted
from item features alone; responses accumulate in real time; and the posterior
narrows until the item is fully calibrated.

The contributions of this paper are: (1) a unified CAT framework that
eliminates pilot testing by treating item parameters as latent variables and
handling cold-start items directly on the operational test; (2) Soft Scoring (S2), a Bayesian scoring rule that attenuates
cold-start items proportionally to their uncertainty; (3) Adaptive Adaptive
Administration (A3), a Thompson sampling policy balancing measurement and
calibration; (4) stochastic Sympson-Hetter exposure control integrated with
the Thompson sampling step; (5) a multidimensional extension for tests with
correlated sub-scores; and (6) empirical validation on the Duolingo English
Test.

\section{Related Work}

\paragraph{Item calibration and piloting.}
Traditional IRT calibration requires hundreds of responses per item
\citep{Lord1980}, collected in a dedicated pilot phase that is slow, costly,
and operationally hazardous: items exposed in practice tests face security
risks \citep{Way1998,LaFlair2022} and motivational confounds that alter
their psychometric properties.
Feature-based IRT models mitigate data sparsity by linking item content to
parameter distributions.
The Linear Logistic Test Model \citep{Fischer1973} pioneered this direction;
more recent work uses NLP features \citep{McCarthy2021} to predict difficulty
and discrimination for new items with no response history.
Even with strong features, however, substantial posterior uncertainty remains,
motivating a fully Bayesian treatment of item parameters at test time.

\paragraph{Bayesian IRT and MCMC calibration.}
\citet{PatzJunker1999} introduced hierarchical MCMC samplers for logistic IRT
models, enabling principled uncertainty quantification over item parameters.
Subsequent work extended these methods to multidimensional models
\citep{Beguin2001} and analyzed the impact of parameter uncertainty on
score estimation \citep{Fox2001}.
General-purpose MCMC is, however, too slow for production CATs with thousands
of items and millions of sessions.
SPICE \citep{nydick2026scalable} addresses this with a scalable parametric
calibration engine that fits latent-regression IRT posteriors offline,
providing the parameter distributions that S2A3 consumes at test time.

\paragraph{CAT item selection.}
Standard CAT selection is greedy maximum-Fisher-information
\citep{Birnbaum1968,Lord1980}: at each step the item maximizing
$F_I(\theta;\,\psi_j)$ at the current ability estimate is chosen, where
$\psi_j$ denotes the item parameters.
For the 2PL model with discrimination $a$ and difficulty $b$,
\begin{equation}
  F_I(\theta; a, b) = a^2\,P(\theta;\,a,b)\,[1-P(\theta;\,a,b)],
  \label{eq:fisher}
\end{equation}
where $P = \sigma(a(\theta-b))$; the 3PL model adds a chance parameter
$c$ so that $P = c + (1-c)\,\sigma(a(\theta-b))$, yielding a different
expression for $F_I$.
Global information methods \citep{Chang1996} select based on KL divergence
over the ability posterior rather than a point estimate, and fully Bayesian
selection \citep{VanderLinden1998} integrates Fisher information over the
posterior.
Shadow testing \citep{VanderLinden2005} further adds content constraints via
integer programming.
All of these methods assume fixed, known item parameters, making them
unsuitable for cold-start items whose parameters are still being learned.

\paragraph{Thompson sampling and exploration--exploitation.}
Thompson sampling \citep{Thompson1933} balances exploration and exploitation
in sequential decisions by drawing parameters from their posteriors and
selecting the action that maximizes expected reward under those draws.
Originally proposed for clinical trials, it was rediscovered for multi-armed
bandits \citep{Chapelle2011,Agrawal2012} where it achieves near-optimal
regret bounds.
In the CAT setting, this naturally unifies measurement (exploitation of
informative items) with calibration (exploration of uncertain ones).

\paragraph{Exposure control.}
Unrestricted information-maximizing selection concentrates administrations on
a few highly informative items, threatening test security
\citep{Way1998}.
The Sympson-Hetter method \citep{SympsonHetter1985} iteratively adjusts
item-specific acceptance probabilities to meet exposure targets.
Conditional exposure control \citep{Linden2000} extends this to stratified
ability groups.
Item cloning \citep{VanderLinden2000} offers a stateless alternative by
sampling uniformly from the top-$k$ most informative items.
S2A3 combines both ideas: Sympson-Hetter weights bound per-item and
per-stratum exposure, while temperature scaling provides stateless
distributional control over the entire selection policy.

\section{The S2A3 Framework: Soft Scoring and Adaptive Administration}
\label{sec:a3}

\subsection{Formal Setup}

At each step $t$, the test-taker's posterior ability
distribution $\pi_t(\theta)$ is initialized to $\mathcal{N}(0,1)$ and updated after each
response.
For each item $j \in \mathcal{I}$, we draw a Thompson sample
$\tilde{\psi}_i \sim p(\psi_j \mid \mathcal{D})$ and compute
\begin{equation}
  r_{t,j} = \mathbb{E}_{\theta \sim \pi_t}\!\bigl[F_I(\theta;\,\tilde{\psi}_j)\bigr],
  \label{eq:reward}
\end{equation}
the expected Fisher information under both ability and item-parameter
uncertainty.
Selecting items that maximize $r_{t,j}$ adds an extra layer of adaptivity
beyond standard CAT---hence \emph{adaptive adaptive administration} (A3).

\subsection{Soft Scoring (S2)}
\label{sec:s2}

Thompson sampling also governs how administered items contribute to the ability
posterior update.
Rather than conditioning on point estimates $\hat{\psi}_i$, the scoring step
integrates over the item's posterior ($\mathcal{D}$ denotes accumulated response data):
\begin{equation}
  \pi_{t+1}(\theta) \;\propto\; \pi_t(\theta)
  \int P(G_t \mid \theta, \psi_j)\, p(\psi_j \mid \mathcal{D})\,d\psi_j.
  \label{eq:soft_score}
\end{equation}
The marginal item response function in Equation~\ref{eq:soft_score} is flatter
than any single-parameter IRF: when parameter uncertainty is high, the marginal
curve has lower effective discrimination, attenuating the item's influence on
$\pi_{t+1}$.
This is the ``soft'' in Soft Scoring---items with greater parameter uncertainty
have less impact on scores (S2), preventing cold-start items from distorting
ability estimates while their parameters are still being learned.

\subsection{The S2A3 Algorithm}

\begin{algorithm}[t]
\caption{S2A3 Item Administration}
\label{alg:s2a3}
\begin{algorithmic}[1]
\State Init $\pi_1(\theta) = \mathcal{N}(0,1)$;\; eligible pool $\mathcal{I}$;\; precompute weights $w_j$
\For{$t = 1, \ldots, T$}
    \For{each $j \in \mathcal{I}$}
        \State Draw $\tilde{\psi}_i \sim p(\psi_j \mid \mathcal{D})$ \hfill \textit{(Thompson sample)}
        \State Compute $r_{t,j} = \mathbb{E}_{\theta \sim \pi_t}[F_I(\theta;\,\tilde{\psi}_i)]$
    \EndFor
    \State Set $p_{t,j} \propto w_j \cdot r_{t,j}^{\beta}$ for all $j \in \mathcal{I}$
    \State Sample $I_t \sim (p_{t,j})$;\; administer $I_t$;\; remove from $\mathcal{I}$
    \State Observe grade $G_t$;\; update $\pi_{t+1}$ via Eq.~\ref{eq:soft_score}
\EndFor
\end{algorithmic}
\end{algorithm}

Items with high posterior variance occasionally yield large $r_{t,j}$ draws
and are selected even when their mean information is modest; as responses
accumulate the posterior sharpens and the policy transitions smoothly to
pure exploitation.

\subsection{Multidimensional Extension}
\label{sec:multidim}

Many operational CATs administer multiple item types that target related but
distinct sub-skills---for example, vocabulary knowledge and reading
comprehension.
We extend S2A3 to this setting by modeling ability as a $K$-dimensional vector
$\boldsymbol{\theta} = (\theta_1, \ldots, \theta_K)$ with a multivariate
normal prior $\boldsymbol{\theta} \sim \mathcal{N}(\mu, \Sigma)$.
Each item type $k$ follows its own unidimensional IRT model
$f(G_t \mid I_t, \theta_{k_t})$, so the joint likelihood factorizes
across item types.

Because the posterior $\pi(\boldsymbol{\theta} \mid \mathbf{G}_t, \mathbf{I}_t)$
is non-Gaussian for logistic IRT models, we use a Laplace approximation.
Since the likelihood factorizes, the precision matrix $\Psi_t$ is diagonal,
and each diagonal entry $\Psi_{t,k,k}$ can be estimated independently from
the responses to item type $k$.
We estimate $\Psi_{t,k,k}^{-1}$ via \emph{one-sided moment matching} on the
likelihood values away from the mode, which is robust in data-sparse settings
where the Hessian-based estimate underestimates curvature.
The approximate posterior mean and covariance are then
\begin{align}
  \hat{\boldsymbol{\mu}}_t &= \bigl(\Sigma^{-1} + \Psi_t\bigr)^{-1}
    \bigl(\Sigma^{-1}\mu + \Psi_t\hat{\boldsymbol{\theta}}_t\bigr),
    \label{eq:multivariate_posterior} \\
  \hat{\Sigma}_t &= \bigl(\Sigma^{-1} + \Psi_t\bigr)^{-1}. \notag
\end{align}
This Laplace approximation is used only during test administration to enable real-time scoring; final operational scoring can use a more accurate posterior approximation without the latency constraints of live item selection.
Each component $k$ inherits its marginal posterior
$\pi_t^k = \mathcal{N}(\hat{\mu}_{t,k},\, \hat{\Sigma}_{t,k,k})$,
which is then used as the ability prior for the next item of type $k$ in the
standard A3 selection step.
The cross-correlations encoded in $\Sigma$ allow responses to one item type to
sharpen estimates for other types---the primary benefit of multivariate
scoring over $K$ independent univariate scorers.
Full derivations of the S2, A3, and multidimensional components, including
proofs and implementation details, are given in \citet{Sharpnack2024}.

\section{Stochastic Sympson-Hetter Exposure Control}
\label{sec:exposure}

\subsection{Selection Probabilities and Temperature Scaling}

The S2A3 algorithm combines Thompson sampling with exposure control through a
\emph{stochastic Sympson-Hetter algorithm with temperature scaling}.
After computing the posterior expected information for each item, we sample
items from a probability distribution that favors items with larger information
content, modulated by two exposure-control parameters: a global inverse
temperature $\beta \geq 0$ and item-specific Sympson-Hetter weights
$w_j \in (0,1]$ for each item $j \in \mathcal{I}$ (see Algorithm~\ref{alg:s2a3}, line 7):
\begin{equation}
  p_{t,j} \;\propto\; w_j \cdot r_{t,j}^{\beta}.
  \label{eq:selection}
\end{equation}
Note that if Sympson-Hetter weights are all set to 1, setting $\beta = 0$
gives all items equal chance of being administered, while sending
$\beta \to \infty$ results in deterministically choosing the item with the
highest Fisher information.
Intermediate $\beta$ values produce a smooth trade-off.

Setting $\beta=0$ gives uniform selection; $\beta\to\infty$ recovers
deterministic maximum-information CAT; intermediate values trade measurement
efficiency for bank diversity.
The $w_j$ weights cap per-item exposure regardless of how informative an item
is, while $\beta$ controls the overall dispersion of the selection
distribution.

\subsection{Bank Health Metrics}

We summarize item exposure and bank health using four metrics:
\begin{enumerate}\setlength{\itemsep}{2pt}
  \item \textbf{Maximum exposure}: the highest marginal administration
    probability across all items.
    A ceiling of $\approx$0.20 is a common standard for high-stakes tests.
  \item \textbf{Maximum conditional exposure (MCE)}: the highest exposure
    probability within any ability stratum; catches items that are overexposed
    to a specific subpopulation even if their marginal exposure appears
    acceptable.
  \item \textbf{Adjusted effective bank size (AEBS)}: the size of a
    hypothetical bank with equal exposure rates that yields the same entropy as
    the current policy.
    Higher AEBS indicates more uniform use of the bank.
  \item \textbf{Rarely-administered fraction}: the proportion of items
    administered in fewer than 0.1\% of sessions; tracks whether the tail of
    the bank is being neglected.
\end{enumerate}

\subsection{Estimating Sympson-Hetter Weights}

The Sympson-Hetter weights $w_j$ and the inverse temperature $\beta$ must be
tuned to achieve target exposure metrics before deploying the algorithm.
We obtain these weights via simulation and by optimizing an
exposure-constrained Fisher information objective.
For any given inverse temperature $\beta$, Sympson-Hetter weights can be found
iteratively to achieve target values of maximum exposure and MCE.
The inverse temperature $\beta$ can then be tuned in an outer loop to achieve
target values for AEBS and the fraction of rarely administered items.
Because the four metrics respond differently to the two parameters---$w_j$
controls the upper tail while $\beta$ controls overall dispersion---the
two-loop structure converges reliably without requiring joint optimization.

\section{Application to the Duolingo English Test}
\label{sec:det}

\paragraph{Item types.}
We apply S2A3 to two vocabulary-focused item types on the Duolingo English Test
(DET) \citep{Cardwell2022}.
\emph{Yes/No Vocabulary} (Y/N Vocab) asks the test-taker to identify whether a
word is real or not; the binary forced-choice format is modeled with a 3PL IRT
model with fixed chance parameter $c = 0.25$.
Each session includes 18 Y/N Vocab items with a 5-second time limit.
\emph{Vocabulary-in-Context} (ViC) presents a masked passage and asks
the test-taker to complete the word; random guessing is unlikely, so a 2PL
model ($c = 0$) is used.
Each session includes 9 ViC items with a 20-second time limit.
Item parameters are estimated by SPICE \citep{nydick2026scalable}, a fully Bayesian
calibration engine that links item features---word frequency, CEFR level, and
neural embeddings---to IRT parameter distributions, enabling predictions for
items with no response history.

\paragraph{Multidimensional administration.}
We model ability as a two-dimensional vector
$\boldsymbol{\theta} = (\theta_{\text{Y/N}},\, \theta_{\text{ViC}})$ with a
bivariate normal prior with correlation $\rho = 0.87$.
Because Y/N Vocab items are administered first, we use the univariate S2A3
algorithm to update $\pi(\theta_{\text{Y/N}})$.
For ViC items, we jointly model
$\boldsymbol{\theta}$ using the multidimensional S2A3 algorithm.
As an additional step for Y/N Vocab, we randomly choose (with 50\% probability)
to administer a real or fake word, then restrict the eligible items to only
that group.

\paragraph{Simulation study.}
SPICE is calibrated on a subset of the operational item bank; true item
parameters are then drawn from the resulting posterior and used to simulate
100{,}000 test sessions.
We compare \textbf{S2A3} (Thompson sampling) against \textbf{Degenerate S2A3}
(posterior-mean point estimates, no exploration), both using the same
Sympson-Hetter exposure-control procedure.

\section{Results}
\label{sec:results}

\subsection{Exposure Control}

We simulated 100{,}000 test sessions with target maximum exposure rate 1\% and
target maximum conditional exposure rate (MCE) 10\%.
Both variants held maximum conditional exposure well within the 10\% target
(5.86--5.98\% for ViC, 3.47--3.87\% for Y/N Vocab).
Maximum marginal exposure was slightly above the 1\% target for both variants
(1.09--1.24\%), indicating that the Sympson-Hetter weights provide
approximate rather than exact enforcement of the marginal constraint.
Adjusted effective bank size (AEBS) was 294--302 across variants, indicating
comparable overall bank utilization.

\subsection{Score Reliability}

Table~\ref{tab:ability} compares ability-estimation accuracy for univariate
and multivariate scoring.

\begin{table}[t]
\centering
\small
\caption{Ability-estimation accuracy from 100K simulated sessions.
Uni.\ = univariate; Multi.\ = multivariate.}
\label{tab:ability}
\begin{tabular}{llcccc}
\toprule
 & & \multicolumn{2}{c}{S2A3} & \multicolumn{2}{c}{Degenerate} \\
\cmidrule(lr){3-4}\cmidrule(lr){5-6}
Mode & Metric & Y/N & ViC & Y/N & ViC \\
\midrule
\multirow{2}{*}{Uni.}
  & Correlation & 0.831 & 0.886 & 0.830 & 0.889 \\
  & RMSE        & 0.556 & 0.463 & 0.559 & 0.458 \\
\midrule
\multirow{2}{*}{Multi.}
  & Correlation & 0.884 & 0.902 & 0.884 & 0.903 \\
  & RMSE        & 0.468 & 0.432 & 0.468 & 0.430 \\
\bottomrule
\end{tabular}
\end{table}

All correlations exceed 0.83, demonstrating reliable ability estimation within
a 4.5-minute window (18 Y/N Vocab items at 5\,s each; 9 ViC items at 20\,s
each).
Multivariate scoring consistently outperforms univariate scoring, reducing RMSE
by 6--16\% and improving correlations by 1.6--6.5\%.
Critically, S2A3 matches the degenerate variant on every accuracy metric,
confirming that the exploration introduced by Thompson sampling does not degrade
scoring reliability.

\subsection{Exploration}

\begin{figure*}[t]
\centering
\includegraphics[width=\textwidth]{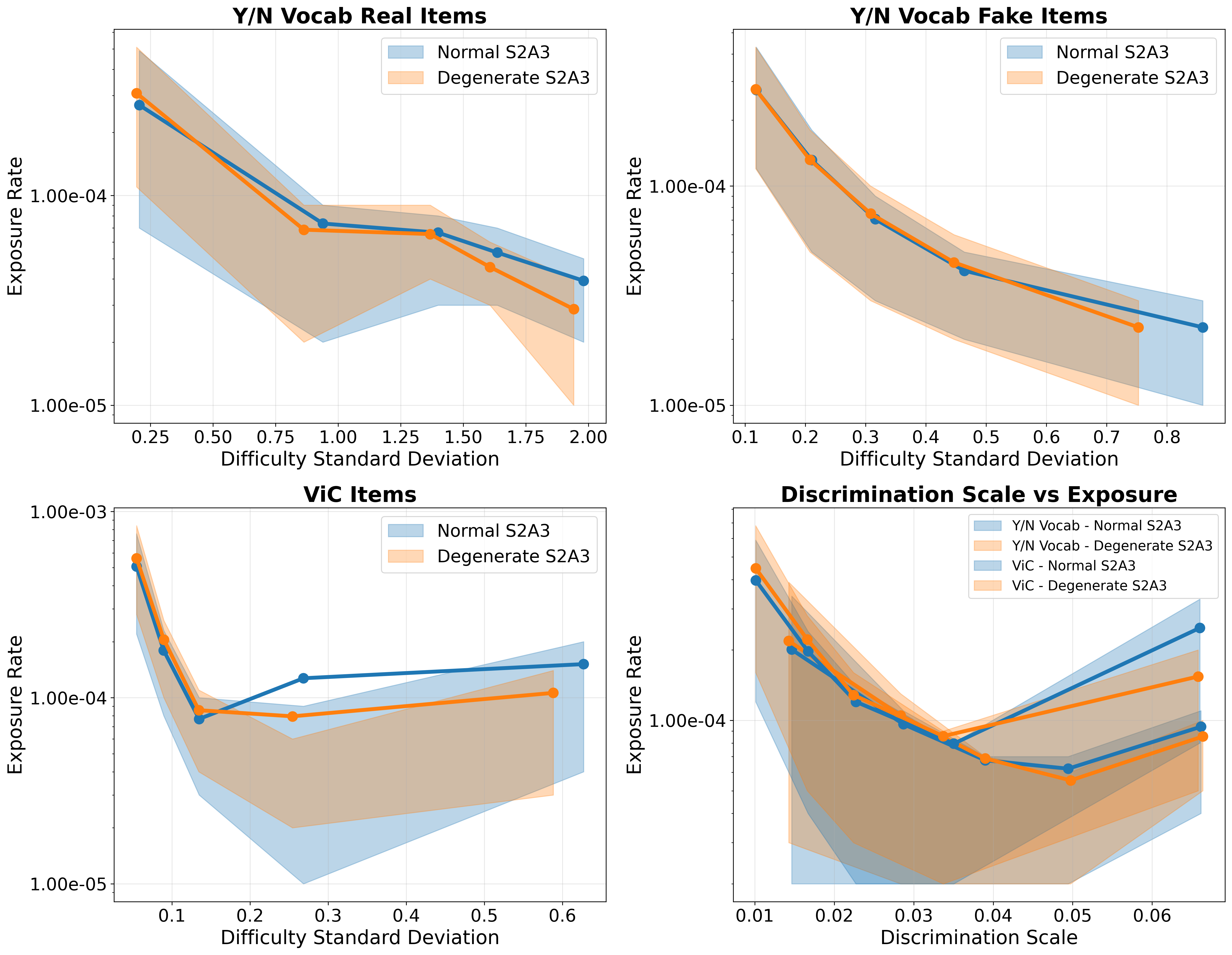}
\caption{Geometric mean exposure rate vs.\ item-parameter variability (posterior
standard deviation of difficulty) for S2A3 and Degenerate S2A3 on ViC items.
Shaded bands show the 25th--75th percentile range.
S2A3 assigns higher exposure to high-uncertainty ViC items.}
\label{fig:exploration}
\end{figure*}

Figure~\ref{fig:exploration} shows that for ViC items, S2A3 assigns
higher exposure to items with larger difficulty or discrimination posterior
standard deviation compared to the degenerate variant.
This effect is most pronounced at the high-uncertainty tail, where Thompson
sampling draws occasionally large parameter values and boosts such items'
expected Fisher information, routing them to more test-takers.
The Y/N Vocab items show a weaker separation because they are administered
first, when the ability posterior is still broad and many items are similarly
informative regardless of their parameter uncertainty.
This confirms that Thompson sampling actively exploits parameter uncertainty
to direct calibration effort toward items that need it most---the core
benefit of A3---without sacrificing the exposure-control guarantees of
Section~\ref{sec:exposure}.

\section{Conclusion}

S2A3 demonstrates that the calibration--administration divide in high-stakes
CAT is not a structural necessity but a consequence of treating item parameters
as fixed unknowns rather than evolving posteriors.
By embedding Thompson sampling in a fully Bayesian policy, every operational
session becomes a calibration event, eliminating the pilot phase entirely.

\paragraph{Empirical summary.}
In 100K simulated DET sessions, S2A3 held maximum conditional exposure to
$\approx$6\% (target: 10\%) and maximum marginal exposure to 1.1--1.2\%,
slightly above the 1\% target, indicating the Sympson-Hetter weights provide
approximate enforcement of the marginal constraint.
Adjusted effective bank size was 294--302 (theoretical maximum: 595),
comparable between S2A3 and the degenerate baseline.
Ability estimation achieved correlations of 0.83--0.90 against ground-truth
abilities within a 4.5-minute test window.
Multivariate scoring reduced RMSE by 6--16\% and improved correlations by
1.6--6.5\% over univariate baselines, confirming that the cross-correlation
between Y/N Vocab and ViC ability ($\rho = 0.87$) provides meaningful
information transfer.
Crucially, full S2A3 (Thompson sampling) matched degenerate S2A3 (point
estimates) on every accuracy and exposure metric, while
Figure~\ref{fig:exploration} confirms that it routes more exposure to
high-uncertainty ViC items---the core calibration benefit of A3.

\paragraph{Automatic bank maintenance.}
A byproduct of Fisher information selection is self-regulating bank quality:
as items accumulate responses and their posteriors sharpen, items with low
posterior-mean discrimination are naturally deprioritized relative to newly
launched, uncertain candidates.
This creates a continuous quality filter without manual curation or explicit
item retirement rules.

\paragraph{Limitations and future work.}
The current implementation is restricted to dichotomous items under the 2PL
and 3PL models.
Polytomous items (e.g., c-test fill-in-the-blank) require a generalized Fisher
information objective and a polytomous soft-scoring integral over the graded
response or partial-credit model likelihood.
Continuous-grade items such as automated speech scoring tasks require a
continuous-grade IRT generalization and a matching reward function.
On the administration side, the item-parameter posterior is currently
approximated by a parametric Normal-Gamma fit to offline MCMC samples;
replacing this with sequential Monte Carlo would enable continuous online
recalibration, removing the need for periodic batch re-runs of SPICE.
Finally, enriching the ability state with auxiliary signals---response latency,
keystroke dynamics, or self-reported confidence---within a richer contextual
bandit model is a promising direction for improving both measurement precision
and examinee experience.
More broadly, S2A3 provides a template for any domain where item banks must be
continuously refreshed, from clinical assessment to professional licensing.

\bibliography{references}

\end{document}